\documentclass[RNAAS,twocolumn]{aastex631}
\usepackage{newtxtext}
\usepackage[slantedGreek]{newtxmath}
\makeatletter
\long\def\frontmatter@title@above{
\vspace*{-\headsep}\vspace*{\headheight}
\footnotesize\noindent{\sc \today}\\[2pt]
{\footnotesize Typeset using \LaTeX\ {\bf RNAAS} style in AASTeX631}
\par\vspace*{-\baselineskip}\vspace*{0.625in}}
\makeatother
\begin{document}

\title{Arcminute Microkelvin Imager observations at 15.5~GHz of multiple
outbursts of Cygnus X-3 in 2024}

\author[0000-0003-3189-9998]{David A.\ Green}
\affiliation{Cavendish Laboratory, University of Cambridge,\\
   19 J.~J.\ Thomson Ave., Cambridge, CB3 0HE, UK}

\author{Lauren Rhodes}
\altaffiliation{current address: McGill University.}
\affiliation{Astrophysics, Department of Physics, University of Oxford,\\
   Denys Wilkinson Building, Keble Road, Oxford OX1 3RH, UK}

\author{Joe Bright}
\affiliation{Astrophysics, Department of Physics, University of Oxford,\\
   Denys Wilkinson Building, Keble Road, Oxford OX1 3RH, UK}

\correspondingauthor{David A.\ Green}
\email{dag@mrao.cam.ac.uk}

\begin{abstract}
We report radio monitoring of \object{Cygnus X-3} at 15.5~GHz during
2024 with the Arcminute Microkelvin Imager. Observations were made on
296 days throughout the year, and reveal five radio outbursts to
multi-jansky levels, peaking in Feb, Apr, Jun, Jul and Aug. The
brightest peak, with $\approx 16$~Jy, was on Jun 27th.
\end{abstract}


\section{}

\object{Cygnus X-3} is a high mass X-ray binary which shows occasional
giant flares at radio wavelengths, to flux densities of up to $\sim
10$~Jy, or more. The first such burst was observed in 1972
\citep{1972Natur.239..440G}, and there have been mulitple similar flares
since (e.g.\ see \citealt{1986ApJ...309..707J, 1995AJ....110..290W,
1997MNRAS.288..849F, 2001ApJ...553..766M, 2012MNRAS.421.2947C,
2016MNRAS.456..775Z, 2017MNRAS.471.2703E} and
\citealt{2020RNAAS...4...36G} for further examples).

We report radio observations of Cygnus X-3 made during 2024 with the
Arcminute Microkelvin Imager (AMI, \citealt{2008MNRAS.391.1545Z,
2018MNRAS.475.5677H}). The observations were made with the AMI `Large
Array' which is a radio interferometer consisting of eight 12.8-m
diameter antennas. A single linear polarisation, Stokes parameter $I+Q$,
was observed over a frequency range of 13 to 18~GHz.

\begin{figure*}[t]
\centerline{\includegraphics[clip=,angle=270,width=16cm]{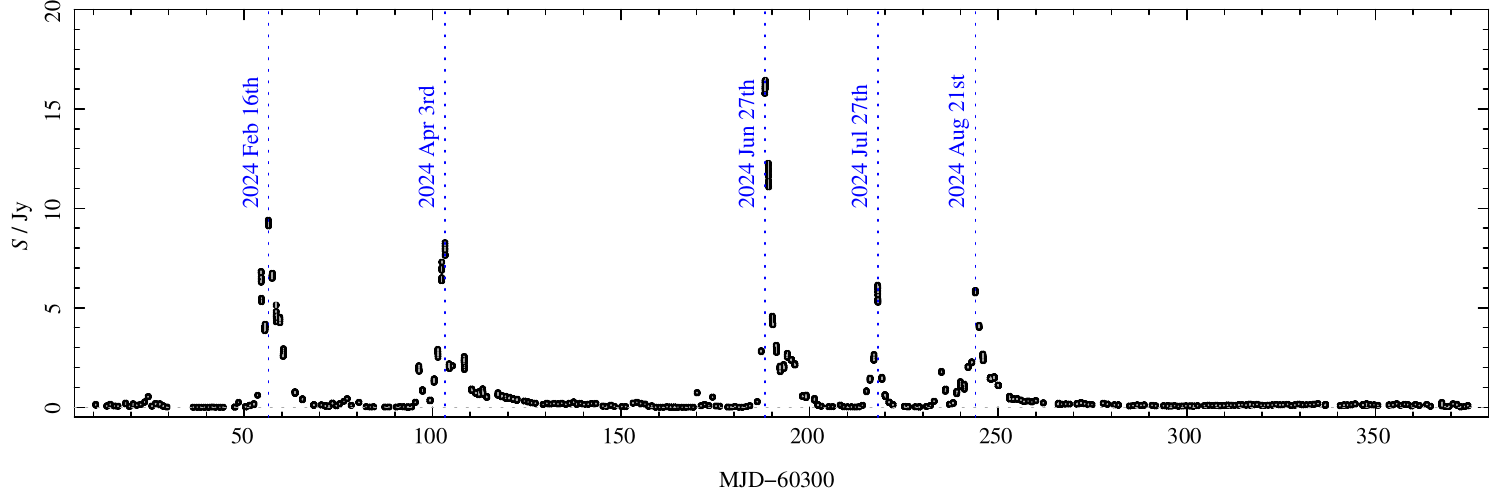}}
\bigskip
\centerline{\includegraphics[clip=,angle=270,width=16cm]{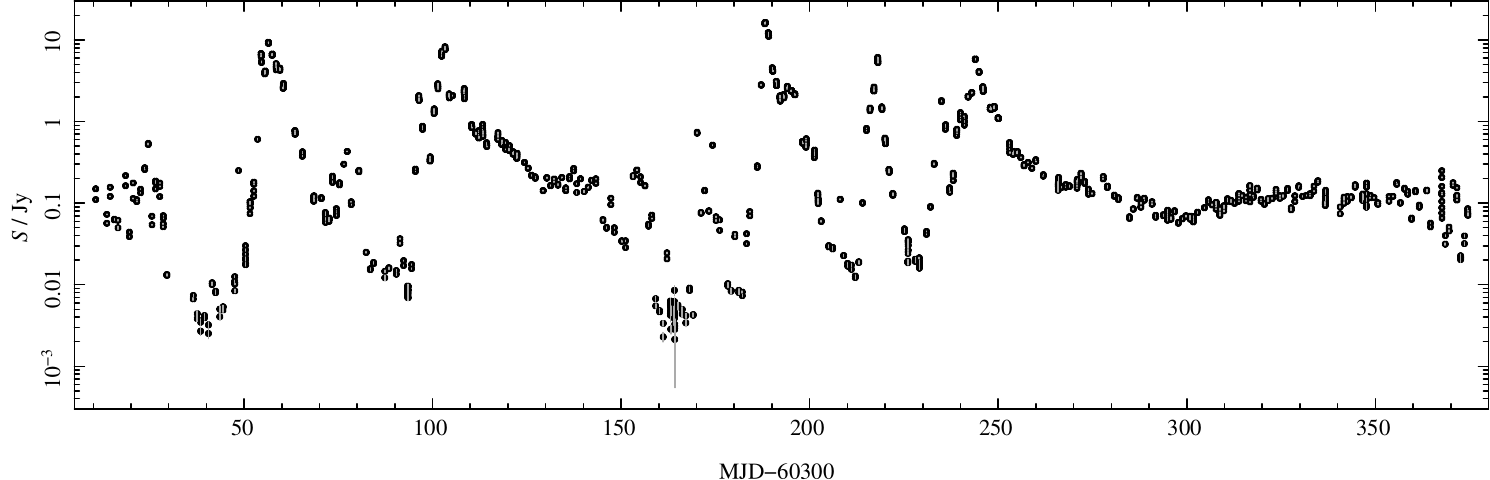}}
\bigskip
\caption{Radio light curve of Cygnus X-3 at 15.5~GHz during 2024 from
AMI observations. The top and bottom panels show, respectively, the flux
density plotted linearly and logarithmically. Each data point is a
10-min average. Statistical error bars are plotted, although these are
usually smaller than the size of symbols. The vertical lines on the top
panel show the dates of the peaks of the outbursts.\label{fig:ami-la}}
\end{figure*}

The observations consisted of multiple 10-min scans of Cygnus X-3,
interleaved with 100-s observations of a nearby, compact calibrator
source J2052$+$3635. Usually an observations consisted of two 10-min
scans of Cygnus X-3, but longer observations were scheduled when the
source was undergoing an outburst, or when other observations were being
made (e.g.\ in X-rays with IXPE or XRISM, for example see
\citealt{2024A&A...688L..27V}). Observations were made on most days in
2024, with omissions due to bad weather (either high winds requiring the
antennas to be stowed, or heavy rain, which makes accurate calibration
not possible), or technical issues. The data were processed using
standard procedures, using the \texttt{reduce\_dc} package (e.g.\
\citeauthor{2018MNRAS.475.5677H}). The flux density scale was
established from short observations of the standard calibrator source
3C286 which were made on most days, together with the `rain gauge'
measurements made during the observations which were use to correct for
varying atmospheric conditions (see \citeauthor{2008MNRAS.391.1545Z}).
The data were flagged: (i) automatically to eliminate bad data due to
various technical problems and interference; (ii) manually, to eliminate
remaining interference and some periods with heavy rain. The interleaved
observations of J2052$+$3635 provided the phase calibration for each
antenna in the array throughout each observation. The amplitudes of the
J2052$+$3635 observations were also used to check and adjust the
day-to-day flux density scale. The day-to-day flux density scale
variations are thought to be less than 5~per~cent. When Cygnus X-3 was
bright during a flare (above 0.3~Jy), the observations were phase
self-calibrated on a timescale of 10 min. Flux densities were derived
for each 10-min scan, for 8 broad frequency channels covering 13 to
18~GHz, and then a power law fit was made to obtain a flux density at
15.5~GHz.

Figure~\ref{fig:ami-la} shows the 15.5-GHz light curve of Cygnus X-3
from these observations. This shows the results from 296 observations
made in 2024, with flux densities for 1407 10-min scans. This shows five
large outbursts, with flux densities $>5$~Jy, which peaked on 2024 Feb
16th, Apr 3rd, Jun 27th, Jul 27th and Aug 21st. During 2022 and 2023
similar monitoring observations of Cygnus X-3 with AMI did not detect
any large outbursts, with the emission typically being at $\sim 0.1$~Jy.
The 2024 observations of Cygnus X-3 shown in Fig.~\ref{fig:ami-la} shows
that the outbursts are preceded by periods when the emission is fainter
than the usual level of $\sim 0.1$~Jy for several weeks, as has been
seen previously (e.g.\ \citealt{2016MNRAS.456..775Z}).

For most of the last three months of the 2024 Cygnus X-3 has shown
little variability at 15.5~GHz from day-to-day. However, towards the end
of the year did show more variation day-to-day. Also, during the
$\approx 1.8$~h observation made on Dec 23 Cygnus X-3 showed clear
variation. It started with $\approx 0.10$~Jy, brightened to $\approx
0.26$~Jy over the next $\approx 0.4$~h, and then faded to $\approx
0.07$~Jy.

\let\internallinenumbers=\relax

\begin{acknowledgments}
We thank the staff of the Mullard Radio Astronomy Observatory,
University of Cambridge, for their support in the maintenance, and
operation of AMI. We acknowledge support from the European Research
Council under grant ERC-2012-StG-307215 LODESTONE.
\end{acknowledgments}

\makeatletter
\newcommand\rnaas{\ref@jnl{RNAAS}}%
\makeatother

\end{document}